\documentclass{emulateapj}
 \usepackage{url}

\shorttitle{GRB 110731A: Lorentz factor, jet composition, central engine, and
progenitor}
\shortauthors{L\"{u} et al}
\slugcomment{ }

\begin{document}

\title{The peculiar GRB 110731A: Lorentz factor, jet composition, central engine, and progenitor}

\author{HouJun L\"{u}\altaffilmark{1}, XiangGao Wang\altaffilmark{1,2}, RuiJing
Lu\altaffilmark{1}, Lin Lan\altaffilmark{1}, He
Gao\altaffilmark{3}, EnWei Liang\altaffilmark{1}, Melissa L.
Graham\altaffilmark{2,7}, WeiKang Zheng\altaffilmark{2}, Alexei
V. Filippenko\altaffilmark{2}, and Bing
Zhang\altaffilmark{4,5,6}} \altaffiltext{1}{Guangxi Key
Laboratory for Relativistic Astrophysics, Department of
Physics, Guangxi University, Nanning 530004, China;
lhj@gxu.edu.cn; lew@gxu.edu.cn} \altaffiltext{2}{Department of
Astronomy, University of California, Berkeley, CA 94720-3411,
USA} \altaffiltext{3}{Department of Astronomy, Beijing Normal
University, Beijing 100875, China}\altaffiltext{4}{Department
of Physics and Astronomy, University of Nevada Las Vegas, Las
Vegas, NV 89154, USA;
zhang@physics.unlv.edu}\altaffiltext{5}{Department of
Astronomy, School of Physics, Peking University, Beijing
100871, China}\altaffiltext{6}{Kavli Institute of Astronomy and
Astrophysics, Peking University, Beijing 100871,
China}\altaffiltext{7}{Astronomy Department, University of
Washington, Box 351580, U.W., Seattle, WA 98195-1580}

\begin{abstract}
The jet compositions, central engines, and progenitors of
gamma-ray bursts (GRBs) remain open questions in GRB physics.
Applying broadband observations, including GRB prompt emission
and afterglow properties derived from {\em Fermi} and {\em
Swift} data, as well as from Keck host-galaxy observations, we
address these questions for the peculiar, bright GRB 110731A.
By using the pair-opacity method, we derive $\Gamma_{0}>190$
during the prompt emission phase. Alternatively, we derive
$\Gamma_{0} \approx 580$ and $\Gamma_{0} \approx 154$ by
invoking the early-afterglow phase within the homogeneous
density and wind cases, respectively. On the other hand,
nondetection of a thermal component in the spectra suggests
that the prompt emission is likely powered by dissipation of a
Poynting-flux-dominated jet leading to synchrotron radiation in
an optically thin region. The nondetection of a jet break in
the X-ray and optical bands allows us to place a lower limit on
the jet opening angle $\theta_j>5.5^{\circ}$. Within a
millisecond magnetar central engine scenario, we derive the
period $P_{0}$ and polar magnetic field strength $B_{\rm p}$, which
have extreme (but still allowed) values. The moderately short
observed duration (7.3\,s) and relatively large redshift
($z=2.83$) place the burst as a ``rest-frame short'' GRB, so
the progenitor of the burst is subject to debate. Its
relatively large $f_{{\rm eff}, z}$ parameter (ratio of the
1\,s peak flux of a pseudo-GRB and the background flux) and a
large physical offset from a potential host galaxy suggest that
the progenitor of GRB 110731A may be a compact-star merger.
\end{abstract}


\section{Introduction}\label{sec:intro}

Despite decades of investigation, there still exist several
open questions in gamma-ray burst (GRB) physics, particularly
regarding their progenitors, central engines, and jet
compositions (e.g., Zhang 2011; Kumar \& Zhang 2015).

Traditionally, GRBs are classified into long/soft and
short/hard categories based on their distributions in the
$T_{90}$ vs. hardness-ratio plane (Kouveliotou et al. 1993).
However, the measurement of $T_{90}$ is energy and
instrument dependent (Qin et al. 2013). Observations of the
host-galaxy properties and supernova associations of GRBs
suggest that the majority of long-duration GRBs originate from
core collapse of massive stars (``collapsars''; Woosley 1993),
while the majority of short-duration GRBs originate from
coalescence of two compact stars (Pacz\'{y}nski 1986; Eichler
et al. 1989). But with only duration information, the physical
category of a GRB is sometimes misclassified (e.g., Gehrels et
al. 2006; Levesque et al. 2010); multiwavelength data are
needed to make correct classifications (Zhang et al. 2009).
Correctly identifying compact-star merger systems is of great
interest, since they are promising gravitational wave sources
to be detected by advanced LIGO/Virgo (e.g., Chu et al. 2016).

Within both the collapsar and compact-star merger models,
another interesting question is what central engine launches
the relativistic outflow. A widely discussed scenario invokes a
hyperaccreting stellar-mass black hole with an accretion rate
of 0.1-1\,$M_{\odot}$\,s$^{-1}$ (e.g., Popham et al. 1999;
Narayan et al. 2001; Chen \& Beloborodov 2007; Liu et al. 2008;
Kumar et al. 2008; Lei et al. 2013). On the other hand, some
GRBs, both long and short, have been discovered to have a
plateau emission component in their X-ray afterglows (Zhang et
al. 2006; O'Brien et al. 2006; Liang et al. 2007), some of them
having an extremely steep drop following the plateau (known as
internal plateaus; Troja et al. 2007; Liang et al. 2007; Lyons
et al. 2010; Rawlinson et al. 2010, 2013; L\"u \& Zhang 2014;
L\"u et al. 2015). Such behavior is consistent with a
millisecond magnetar central engine (Usov 1992; Dai \& Lu 1998;
Zhang \& M\'{e}sz\'{a}ros 2001; Gao \& Fan 2006; Fan \& Xu
2006; Metzger et al. 2010). The steep drop at the end of the
plateau may be consistent with the collapse of a supramassive
neutron star (NS) to a black hole (e.g., Zhang 2014), which has
profound implications for the inferred NS equation of state
(Fan et al. 2013a; Lasky et al. 2014; Ravi \& Lasky 2014; L\"u
et al. 2015; Gao et al. 2016). The existence of a supramassive
NS as a compact-star merger product also enhances
electromagnetic signals of gravitational wave sources (Zhang
2013; Gao et al. 2013a; Yu et al. 2013; Metzger \& Piro 2014;
Fan et al. 2013b), which gives encouraging prospects of
confirming the astrophysical nature of gravitational wave
sources detected by advanced LIGO/Virgo (e.g., Abbott et al.
2016). One challenging task is to distinguish millisecond
magnetars (possibly supramassive) from hyperaccreting black
holes based on data.

The next open question is regarding the composition of the
relativistic jet launched from the central engine, as well as
how energy is dissipated to give rise to prompt emission.
Competing models include the fireball internal shock model
(Rees \& M\'esz\'aros 1994), dissipative photosphere models
(Thompson 1994; Rees \& M\'esz\'aros 2005; Pe'er et al. 2006),
and the internal-collision-induced magnetic reconnection and
turbulence (ICMART) model (Zhang \& Yan 2011; Deng et al.
2015). The first two models have the magnetization parameter
$\sigma$ much less than unity at the GRB emission site, while
the ICMART model has a moderately large $\sigma>1$ at the
emission site, with the GRB emission powered by directly
dissipating the magnetic energy to radiation. These models have
distinct predictions for GRB spectra, light curves, and other
properties. Observations can be used to differentiate among
them.

Finally, it is well known that most of the broadband afterglow
emission is produced from forward and reverse external shocks
(M\'{e}sz\'{a}ros \& Rees 1997; Sari et al. 1998; Kobayashi
2000; M\'{e}sz\'{a}ros 2002; Zhang \& M\'{e}sz\'{a}ros 2004;
Gao et al. 2013b), but the properties of the ambient medium as
well as the shock microphysics parameters remain poorly
constrained (Santana et al. 2014; Wang et al. 2015).

GRB 110731A is a bright GRB jointly detected by the {\em Swift}
and {\em Fermi} satellites. The abundant data collected from
the burst make it a good target to address the open questions
discussed above (e.g., Ackermann et al. 2013; Lemoine et al.
2013; Fraija 2015; Hascoet et al. 2015). In this paper, we
reduce the available high-energy data and present new
observations of the host galaxy of the GRB (\S2). We use the
broadband data to constrain the properties of the GRB,
including its bulk Lorentz factor (\S3), jet composition (\S4),
central engine (\S5), and progenitor (\S6). Our conclusions and
discussion are given in \S7.

\section{Data Reduction and Analysis}
\subsection{Data Reduction}

Both the {\em Swift} Burst Alert Telescope (BAT) and the {\em
Fermi} Gamma-ray Burst Monitor (GBM) triggered GRB 110731A. The
burst was also detected by the Large Area Telescope (LAT), so
high-energy photons above 100\,MeV were detected with
$>10\sigma$ significance (Ackermann et al. 2013). The {\em
Swift} X-Ray Telescope (XRT) and Ultraviolet/Optical Telescope
(UVOT) promptly slewed to the source 56\,s and 62\,s after the
BAT trigger, respectively. Bright X-ray and optical afterglows
were detected with a spectroscopic redshift $z=2.83$ identified
(Tanvir et al. 2011).

\subsection{Prompt Emission}

The {\it Fermi} Gamma-ray Space Telescope comprises two science
instruments, the GBM (Meegan et al. 2009) and the LAT (Atwood
et al. 2009). The GBM has 12 sodium iodide (NaI) detectors
covering an energy range from 8\,keV to 1\,MeV, and two bismuth
germanate (BGO) scintillation detectors sensitive to higher
energies between 200\,keV and 40\,MeV (Meegan et al. 2009). The
signals from each of the 14 GBM detectors have three different
types: CTIME, CSPEC, and TTE. The TTE event data files contain
individual photons with time and energy tags. The LAT observes
the energy of photons from 20\,MeV to 300\,GeV (Atwood et al.
2009). The standard LAT analysis is performed with the latest
Pass 8 release data ($>100$\,MeV). However, the LAT Low Energy
(LLE) data are produced by increasing the effective area of the
LAT at low energy ($\sim30$~MeV) and with very loose event
selection. It required only minimal information, such as the
existence of a reconstructed direction. Also, it is suitable
for studying transient phenomena, such as GRBs. The LLE
analysis also has been updated to the latest Pass 8 event
reconstruction; for more information, refer to the official
{\it Fermi}
website\footnote{http://fermi.gsfc.nasa.gov/ssc/data/}. As
suggested by the {\it Fermi}
team,\footnote{http://fermi.gsfc.nasa.gov/ssc/data/analysis/LAT\_caveats.html
}, it is suitable to use standard LAT event data at high
energies and LLE data at low energies.

\begin{figure}
\center
\includegraphics[angle=0,scale=0.35]{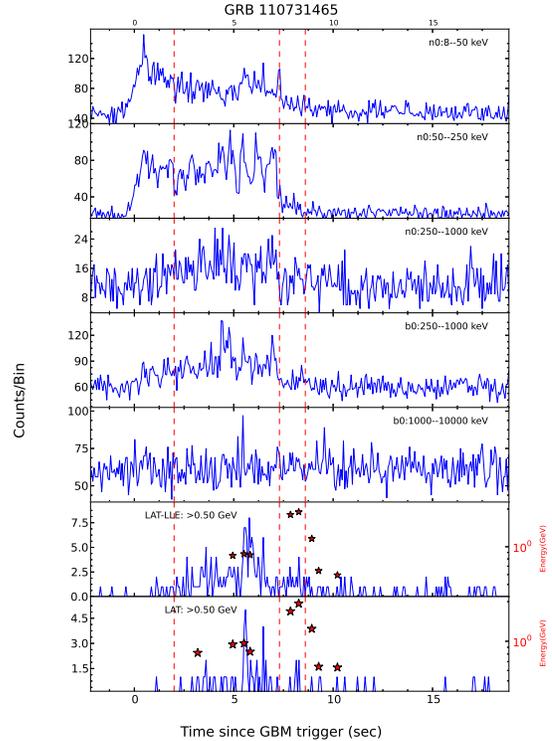}
\hfill\caption{GBM and LAT light curves of GRB 110731A $\gamma$-ray emission in
different energy bands. The stars indicate the LLE and LAT high-energy photons above
500\,MeV. The vertical dotted lines indicate the time intervals for the spectral fitting.}
\label{fig:GBMLC}
\end{figure}

Based on the standard \texttt{heasoft} tools (v. 6.19) and the
{\it Fermi} \texttt{ScienceTool} (v10r0p5), the {\em PYTHON}
source package
$gtBurst$\footnote{http://sourceforge.net/projects/gtburst/} is
designed to analyze the GBM and LAT data, as well as the LLE
data. A step-by-step guide to
the $gtBurst$ can be found on the
website\footnote{fermi.gsfc.nasa.gov/ssc/data/analysis/scitools/gtburst.html/}.
We downloaded GBM, LLE, and LAT data for GRB 110731A from the
public science support center at the {\it Fermi} website. Then,
we extracted the light curves and performed spectral analysis
based on the package {\em gtBurst}. By invoking the
\texttt{heasoft} command \texttt{fselect} and the
\texttt{ScienceTool} command \texttt{gtbin}, we extracted light
curves with a bin size of 0.064~s. However, for the standard
LAT data, we employed an unblinded likelihood analysis method
to build the LAT light curve based on {\em gtBurst}. By
invoking the \texttt{Standardcut} function in $gtBurst$, we
adopted the photons above 100\,MeV in a region of interest
of $12^\circ$ and excluded the photons with zenith angle $>
100^\circ$ to avoid the contribution of Earth's limb. Then, we
extracted the light curve with a bin size of 0.1~s by using the
command \texttt{gtbin}. The resulting light curves of the GBM,
LAT, and LLE data for GRB 110731A are shown in Figure
\ref{fig:GBMLC}. The high-energy photons above 500\,MeV from
the LLE and LAT are also overplotted at the bottom of Figure
\ref{fig:GBMLC}.

\begin{figure}
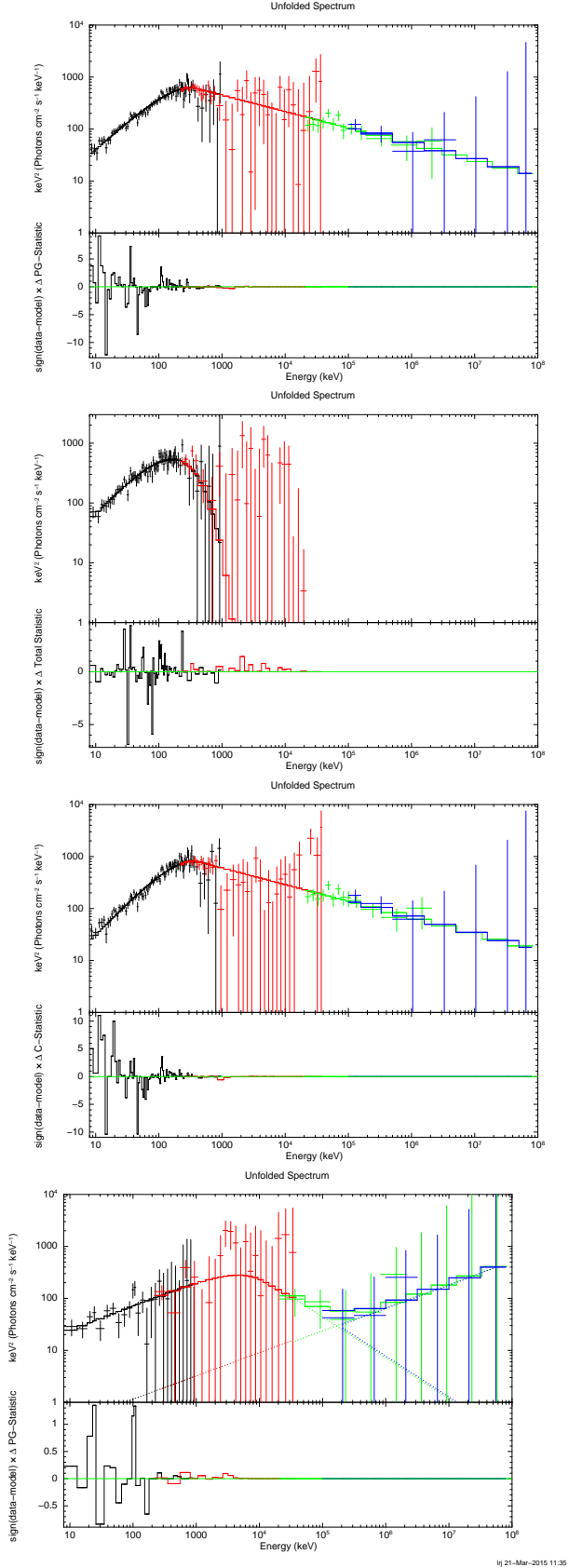

\center
\includegraphics[angle=-90,scale=0.3]{GBMinter.eps}
\includegraphics[angle=-90,scale=0.3]{GBMfirst.eps}
\includegraphics[angle=-90,scale=0.3]{GBMsecond.eps}
\includegraphics[angle=-90,scale=0.3]{GBMthird.eps}
\center\caption{Integrated and time-resolved spectra of GBM-LAT were
fit by a Band function, cutoff power law, and an extra power law. Top to
bottom, the ranges in time are 0--8.6\,s (integrated spectrum), 0--2\,s,
2--7.3\,s, and 7.3--8.6\,s. The black, red, green, and blue points indicate
NaI, BGO, LLE, and LAT data types, respectively.}
\label{fig:GBMSP}
\end{figure}

By invoking ``Tasks$\rightarrow$Make spectra for XSPEC'' in
$gtBurst$, we extracted the source spectra of the GBM and LLE
data. The background spectra are extracted from the time
intervals before and after the prompt emission phase and
modeled with a polynomial function. Then, we extracted the
source spectra by applying the background model to the prompt
emission phase. We derived the LAT spectrum files and response
files by invoking ``Tasks$\rightarrow$Make likelihood
analysis'' in $gtBurst$. First, a standard cut was performed as
done above. The Galactic interstellar emission model
(\texttt{gll\_iem\_v06.fits}) and the isotropic spectral
template (\texttt{iso\_P8R2\_SOURCE\_V6\_V06.txt}) will be used
to reduce the Galactic diffuse and isotropic emission
background contamination. A simple three-component model (an
isotropic diffuse component, a Galactic diffuse component, and
a point source with a power-law spectrum) and the IRF
P8R2\_TRANSIENT020\_V6 were used for unbinned likelihood
analysis. Second, the corresponding response file and
background spectrum file of the GRB's \emph{PHA1} spectrum file
were obtained by using the tools of \texttt{gtbin},
\texttt{gtrspgen}, and \texttt{gtbkg}. Finally, we adopted
XSPEC to conduct a joint spectral fit with a CSTAT statistic
method with the GBM, LLE, and LAT spectra. The results for
three time-resolved spectra and a time-integrated spectrum are
shown in Figure \ref{fig:GBMSP}.

The duration of GRB 110731A is $T_{90}=(7.3\pm0.3)$\,s in the
energy band 50--300\,keV (Gruber 2011), and the fluence is
$\sim (2.22\pm0.01)\times 10^{-5}$\,erg\,cm$^{-2}$ in the
energy band 10--1000\,keV. The time-averaged (from $T_0$\,s to
$T_0+8.6$\,s, where $T_0$ is the BAT trigger time) spectrum is shown
in Figure 2, which can be well fit by a Band function with $E_{\rm
peak}=299^{+54}_{-44}$\,keV,
$\hat{\alpha}=-0.92^{+0.07}_{-0.06}$, and
$\hat{\beta}=-2.34^{+0.02}_{-0.03}$ without the need for an
additional thermal component\footnote{In order to avoid
confusion with temporal ($\alpha$) and spectral ($\beta$)
indices, we use $\hat{\alpha}$ and $\hat{\beta}$ to indicate
the low-energy and high-energy photon spectral indices of the
Band function.}. According to the concordance cosmology with
parameters $H_0 = 71$\,km\,s$^{-1}$\,Mpc$^{-1}$,
$\Omega_M=0.30$, and $\Omega_{\Lambda}=0.70$, the total
isotropic-equivalent energy in the 10\,keV--10\,GeV band is
$E_{\gamma,{\rm iso}}=(6.8\pm 0.1)\times 10^{53}$\,erg with
$z=2.83$.

We also extracted light curves obtained with {\em Swift}/BAT.
For this process, we developed an IDL script to automatically
download and maintain the {\em Swift}/BAT data, and then we used
standard HEASOFT tools (v. 6.12) to process the data. By
running {\em bateconvert} from the HEASOFT software release, we
obtain the energy scale for the BAT events. The light curves
are extracted by running {\em batbinevt} in the 15--150\,keV
energy range (bottom panel in Fig. \ref{fig:BATLAT}). The
time-averaged spectrum can be best fit by a simple power-law
model with photon index $\Gamma_{\rm ph}=1.24\pm 0.08$ owing to
the narrow energy range. No obvious spectral evolution was seen
in the time-resolved spectral BAT data. The spectral index
within the range 6.5--8.6\,s is $\Gamma_{\rm ph}\approx 1.5$.

\subsection{GeV Flare}
\begin{figure}
\center
\includegraphics[angle=0,scale=0.4]{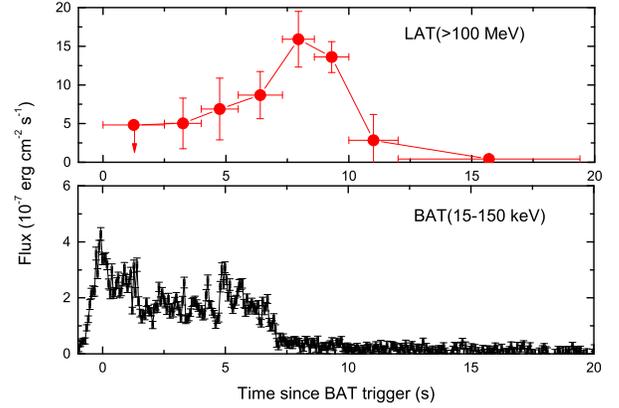}
\center\caption{Top: {\em Fermi}/LAT light curve with energy range
100\,MeV--100\,GeV. Bottom:
{\em Swift}/BAT light curve with energy range 15\,keV--150\,keV.}
\label{fig:BATLAT}
\end{figure}

\begin{figure}
\center
\includegraphics[angle=0,scale=0.4]{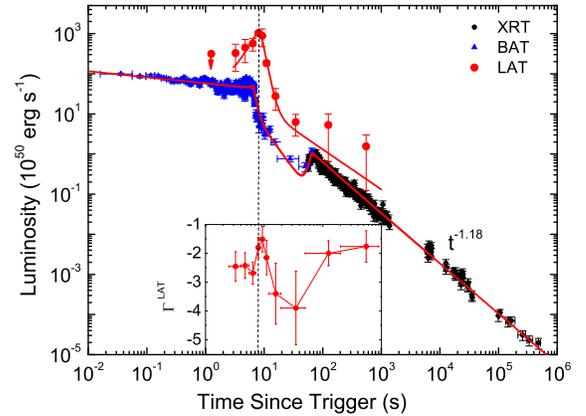}
\caption{Light curves of GRB 110731A observed with {\it Swift}/BAT (blue
triangles; 15\,keV--150\,keV), XRT (black filled circles; 0.3\,keV--10\,keV),
and {\em Fermi}/LAT (red filled circles; 100\,MeV--100\,GeV). The empirical fit with
the power-law model is shown as a solid line. The evolution of the photon
index in the LAT band is also shown in the inset. The vertical line marks
the peak time of the GeV flares observed with LAT.}
\label{fig:BATXRT}
\end{figure}

Figure \ref{fig:BATLAT}) shows the minute-scale structure of
the LAT light curves. We fit the light curves with a model of
multiple power-law components: a power-law function
$F=F_{0}t^{\alpha}$, or a broken power-law function
$F=F_{0}[(t/t_{\rm p})^{\alpha_{1}s}+(t/t_{\rm
p})^{\alpha_{2}s}]^{1/s}$, where $\alpha$, $\alpha_1$, and
$\alpha_2$ are the temporal slopes, $t_{\rm p}$ is the peak
time, and $s$ measures the sharpness of a peak of the
light-curve component. One has a fast-rising ($t^{2.5\pm
0.12}$) phase initially within the time interval 3.2--7.5\,s,
followed by a very rapid decay ($t^{-7.4\pm 0.23}$) phase
within the time interval 7.5--16.5\,s (see Figs.
\ref{fig:BATLAT} and \ref{fig:BATXRT}). Finally, one flat
component follows the rapid decay phase. The peak of the GeV
emission is at around 8\,s. Two highest-energy photons
($\sim2$\,GeV) were detected in the time interval of 8--9\,s
after the BAT trigger (see Fig. \ref{fig:GBMSP}, which shows
the arrival time for high-energy photons above 500\,MeV). The
LAT spectrum is fitted well by a single power-law model,
$N(E)\propto E^{-\Gamma_{\rm LAT}}$, and a clear hard-to-soft
spectral evolution during the decay phase is present.

Ackermann et al. (2013) and Fraija (2015) suggested that this
GeV emission is the afterglow onset from the external shock.
However, the steep temporal indices of both the rising and
decaying phases, as well as the hard-to-soft spectral
evolution, disfavor the afterglow onset scenario. Rather, it is
more like flare emission, which is similar to the X-ray flare
emission commonly observed in X-ray afterglows (Burrows et al.
2005; Chincarini et al. 2007; Margutti et al. 2010). We
therefore define this GeV emission as a GeV flare and discuss
its physical origin. The isotropic energy and luminosity of
this GeV flare are $E_{\rm flare, iso} \approx 1.4\times
10^{53}$\,erg and $L_{\rm flare, iso} \approx 1.1\times
10^{53}$\,erg\,s$^{-1}$ with $z=2.83$. Figure \ref{fig:BATLAT}
shows a comparison between the light curves in the LAT and BAT
bands.

\subsection{X-Ray Afterglow}
We take the XRT data from the {\em Swift} UK XRT team
website\footnote{http://www.swift.ac.uk}. The X-ray afterglow
light curve shows a rapid increase ($t^{7.0\pm0.82}$) and then
a normal decay ($t^{-1.18\pm0.08}$), with a peak time at
$65\pm8$\,s in the observed frame\footnote{Ackermann et al.
(2013) explained the rapid flux increase in the XRT band around
65\,s as an X-ray flare. However, since both the X-ray and
optical light curves after the peak decay as a power-law
function with a typical index, the initial rapid increase of
the XRT light curve would be more like the onset of the
afterglow with flickering contamination, as observed in GRB
061007 (Mundell et al. 2007).}. No jet break feature was
detected up to $\sim 7.5$\,days. The photon index during the
X-ray afterglow phase is $\Gamma_{\rm XRT}\approx 2.0\pm 0.1$,
which is similar to the photon index in the LAT band. Figure 4
shows the BAT and XRT light curves from the GRB trigger to
$\sim 10^6$\,s later. One finds that the prompt emission phase
is essentially a very short plateau with a decay index
$t^{-0.15\pm0.09}$ on a logarithmic scale. Then, it is followed
by a sharp drop $\propto t^{-8.6\pm1.12}$ within the time
interval 6.5--8.6\,s\footnote{However, the temporal decay index
is dependent on the zero-time ($T_{\rm z}$), which is more
difficult to determine. In our calculation, we assume that the
BAT trigger time $T_0$ is also the zero-time.}, and the break
time is $t_{p} \approx 6.5\pm0.8$\,s.

\subsection{Keck Observation of a Potential Host Galaxy}

\begin{figure}
\center
\includegraphics[angle=0,scale=0.27]{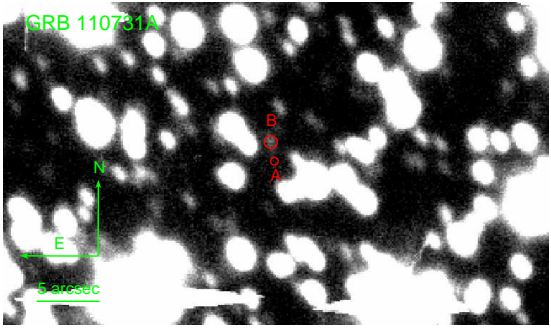}
\caption{Slightly trailed Keck I image of the field of GRB 110731A in the $I$ band. The lower circle (circle A) marks the
position of the GRB 110731A
optical afterglow with $0.3''$ uncertainty and a magnitude limit
$m_{I}\approx 24.9$. The upper circle (circle B) marks the position
of its potential host galaxy, with $0.5''$ uncertainty and
$m_{I}= 23.5$\,mag. The separation between circles A and B is
$1.63''$, which corresponds to $\sim 13.0$\,kpc at $z=2.83$.}
\label{fig:Host}
\end{figure}

We took images of GRB 110731A using the Low Resolution Imaging
Spectrometer (Oke et al. 1995) with the Keck I 10\,m
telescope on 15 June 2015 UT in the $I$ and $V$ filters with $3
\times 400$\,s exposures (see Fig. \ref{fig:Host}). We use the
optical image from GROND to double-check the position of GRB
110731A, which is at (J2000) $\alpha = 18^h42^m01.011^s$,
$\delta = -28^\circ32'13.43''$ (Ackermann et al. 2013). The
foreground stars are very saturated and crowded; thus, in order
to calibrate the magnitudes of stars in the field, we took an
image with the 1.0\,m Nickel telescope at Lick Observatory on
2015 June 30 UT. At the exact afterglow position, we did not
find an apparent host galaxy of GRB 110731A, with a limiting
magnitude of $m_{I}\approx 24.9$.

On the other hand, we found a potential extended source to the
northeast of the afterglow position, which might be the host
galaxy of GRB 110731A. The source has an $I$-band magnitude of
$m_I \approx 23.5$ (as shown in Fig. 5 with circle B).
Following Bloom et al. (2002) and Berger (2010), we calculate
the probability of association for a given galaxy of brightness
$m$ at a separation $\delta R$ from a GRB position:
\begin{eqnarray}
P=1-e^{-\pi(\delta R)^2\sum(\leq m)},
\label{depth}
\end{eqnarray}
where the galaxy number counts are given by $\sum(\leq m)= 1.3
\times 10^{0.33(m-24)-2.44}$\,arcsec$^{-2}$ (Hogg et al. 1997).
Assuming a physical association, the offset
between this putative host galaxy and the GRB afterglow is
$1.63''$, which corresponds to $\sim 13.0$\,kpc at $z=2.83$.
We find that the possibility that GRB 110731A resides in this
host galaxy is about 3\%.

\section{Constraints on the Bulk Lorentz Factor}

The broadband data allow us to constrain the bulk Lorentz
factor $\Gamma$ of the GRB with two different methods.

The first is the pair-opacity method. The broadband
featureless Band-function spectra extending to very high
energies ($>100$\,MeV) pose a lower limit on $\Gamma$ to avoid
two-photon pair production ($\gamma \gamma\rightarrow
e^{+}e^{-}$). If a cutoff energy is detected in the spectrum,
one may constrain $\Gamma$ and the radius of the emission
region ($R_\gamma$) by requiring the optical depth (Zhang \&
Pe'er 2009) to be
\begin{eqnarray}
\tau_{\gamma\gamma}(E_{\rm cut}) & = &
\frac{F(\hat{\beta})\sigma_{T}D^{2}_{L}f_0}{-1-\hat{\beta}}
\left(\frac{E_{\rm cut}}{m^{2}_{e}c^{4}}\right)^{-1-\hat{\beta}} \nonumber \\
& \times & R^{-2}_{\gamma} \left(\frac{\Gamma}{1+z}\right)^{2+2\hat{\beta}} = 1,
\label{depth}
\end{eqnarray}
where $D_{\rm L}$ and $z$ are the luminosity distance and redshift,
respectively, and $m_{e\rm }$, $c$, and $\sigma_{\rm T}$ are the
fundamental constants of electron mass, speed of light, and
Thomson cross section, respectively. The parameter $f_0$ is
related to the Band-function parameters as
\begin{eqnarray}
f_0= A \cdot \Delta T
\left[\frac{E_{\rm p}(\hat{\alpha}-\hat{\beta})}{2+\hat{\alpha}}\right]^{\hat{\alpha}-\hat{\beta}}
\rm
exp(\hat{\beta}-\hat{\alpha})(100\,{\rm keV})^{-\hat{\beta}},
\label{f0}
\end{eqnarray}
where $\Delta T$ is the time interval for spectral fitting,
8.6\,s. Also, $\hat{\alpha}$, $\hat{\beta}$, and $E_{\rm p}$
are the parameters of the time-integrated spectrum within
0--8.6\,s by invoking a Band-function fit, and the fitting
results are presented in Table 1. The Band-function
normalization is $A =
0.044$\,ph\,cm$^{-2}$\,s$^{-1}$\,keV$^{-1}$. The coefficient
$F(\hat{\beta})$ can be expressed as (Svensson 1987)
\begin{eqnarray}
F(\hat{\beta})= \frac{7}{6(1-\hat{\beta})(-\hat{\beta})^{5/3}}.
\label{Fbeta}
\end{eqnarray}

\begin{figure}
\center
\includegraphics[angle=0,scale=0.4]{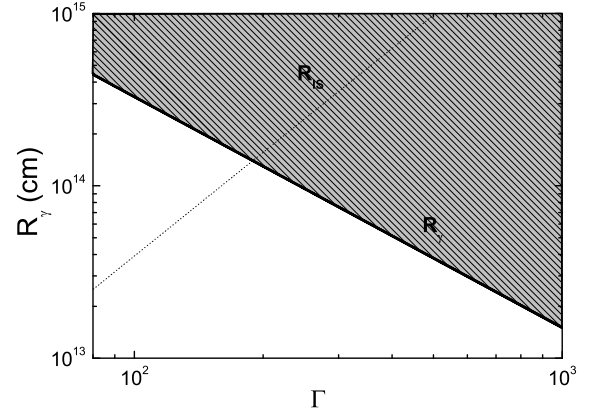}
\hfill\caption{The $R_{\gamma}$--$\Gamma$ diagram of GRB 110731A.
The constraint on $R_{\gamma}$ is displayed with a black solid line,
above which is the allowed parameter space (gray shadow). The internal
shock radius, $R_{\rm IS}\approx \Gamma^{2} c \delta t/(1+z)$,
is plotted as the dotted line.}
\label{fig:Gammar}
\end{figure}

For the spectra of GRB 110731A, a cutoff power-law model is not
consistent with the data. However, a Band function combined
with an extra power-law component provides a good fit from
7.3\,s to 8.6\,s. There is no cutoff feature in the spectra
when this extra power-law component extends to the GeV energy
band. The highest-energy observed photon had $\sim1.9$\,GeV.
Hence, we can use this maximum-energy photon $E_{\rm max}$ to
replace $E_{\rm cut}$ in Eq.~2, and we estimate the lower limit of
the bulk Lorentz factor ($\Gamma_{\rm min}$) by assuming that
both $>100$\,MeV and sub-MeV photons are from the same zone. By
using the $\tau_{\gamma\gamma}(E_{\rm max})\leq 1$ condition,
one derives $\Gamma_{\rm min} \approx 190$ within the
internal-shock model, $R_{\gamma}\approx \Gamma^2 c \delta
t/(1+z)$, where $\delta t$ is the minimum variability
timescale. But $\delta t$ is subject to large uncertainties
because GRB light curves are chaotic, without a characteristic
timescale, and it also depends on the bin size and energy. In
our calculation, $\delta t=0.5$\,s is adopted to get a higher
signal-to-noise ratio for the GBM-LAT light curve. The
constraints on the range of $\Gamma$ and $R_\gamma$ are shown
in Figure \ref{fig:Gammar}. However, if the $>100$\,MeV and
sub-MeV photons are from different regions, then the estimated
bulk Lorentz factor is more complex (Zhao et
al. 2011; Zou et al. 2011).

\begin{figure}
\center
\includegraphics[angle=0,scale=0.4]{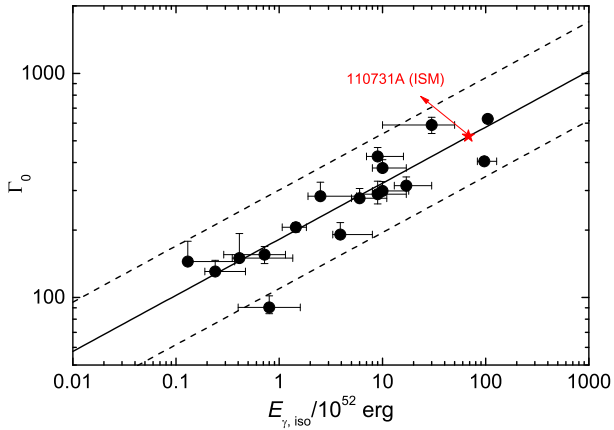}
\hfill\caption{Comparison of GRB 110731A estimated from the early
afterglow onset time with other typical long
GRBs in the $\Gamma_0$--$E_{\gamma, {\rm iso}}$ plane. The filled star
is GRB 110731A assuming a homogeneous density profile (ISM).
GRBs marked with dots are taken from Liang et al. (2010).
Two dashed lines mark the $2\sigma$ region of the correlation, and
$\sigma$ is the standard deviation of the best fit.}
\label{fig:GammaEiso}
\end{figure}

The second method of estimating the initial Lorentz factor is
to use the onset time of the early afterglow. According to a
broken power-law fit, the early X-ray afterglow light curve of
GRB 110731A peaks at $t_{\rm peak} \approx 65$\,s. The
deceleration time is $t_{\rm dec}=t_{\rm peak}/(1+z)$. We apply
the standard afterglow model with a constant-density medium
(i.e., the interstellar medium [ISM]) to derive the initial
Lorentz factor, which reads
\begin{eqnarray}
\Gamma_0 &
\approx & 170\ t_{\rm dec,2}^{-3/8}\left(\frac{1+z}{2}\right)^{3/8}E_{\rm iso,52}^{1/8}
(n\eta)^{-1/8}.
\label{gamma}
\end{eqnarray}
We take $n \approx 0.1$ cm$^{-3}$ and radiative efficiency
$\eta=\frac{E_{\gamma, {\rm iso}}}{E_{\gamma, {\rm iso}}+E_{\rm
K, iso}}\approx 90\%$ in this analysis and derive\footnote{Sari \& Piran (1999) derived $\Gamma_0 =
[{3E_{\rm iso}(1+z)^{3}} / {32\pi n \eta m_p c^5 t_{\rm dec}^3}
]^{1/8}$. L\"u et al. (2012) introduced a factor of 1.4 through
numerical integration. A more precise treatment by including
pressure in the energy-momentum tensor gives our numerical
coefficient.}$\Gamma_0 \approx 580$. Liang et al. (2010) discovered a
tight relation between $\Gamma_0$ and $E_{\gamma, {\rm iso}}$.
We test to compare whether GRB 110731A is consistent with this
correlation. The values $n \approx 1$\,cm$^{-3}$ and $\eta
\approx 20$\% (performed by Liang et al. 2010) are adopted to
recalculate the Lorentz factor of GRB 110731A. One has
$\Gamma_0 \approx 525$, which is consistent with the
correlation between $\Gamma_0$ and $E_{\gamma, {\rm iso}}$
discovered by Liang et al. (2010). Figure \ref{fig:GammaEiso}
shows the $\Gamma_0$--$E_{\gamma, {\rm iso}}$ plot and
compares GRB 110731A in a constant-density ISM (filled
star) with other typical long GRBs taken from Liang et al
(2010). Alternatively, if the medium has a density that decays
with radius (wind model), then
\begin{eqnarray}
\Gamma_0 &
= & 1.44[\frac{E_{\rm iso}(1+z)}{8\pi A_{\ast}m_p c^{3}\eta t_{\rm dec}}]^{1/4}
\nonumber \\
&\approx& 40(\eta E_{\rm iso, 52})^{1/4}(\frac{1+z}{2})^{1/4}(t_{\rm dec, 2})^{-1/4},
\label{gamma}
\end{eqnarray}
where $A_{\ast}=3\times 10^{35}$ cm$^{-1}$ is the wind
parameter (Yi et al. 2015) and $\eta \approx 90$\% is adopted.
One has $\Gamma_0\approx 154$, which is much lower than in the
ISM case.

\section{Jet Composition}

Different jet energy dissipation models of prompt emission
predict different properties in the emission region, such as
the magnetization parameter $\sigma$, the spectral shape, and
the $R_{\gamma}$--$\Gamma$ relation. As shown in Figure 6, the
$\tau_{\gamma\gamma}(E_{\rm max})\leq 1$ condition places a
tight lower limit on $R_\gamma$ for the burst, which is much
larger than the photosphere radius (typically at $R_{\rm
ph}\approx 10^{11}$--$10^{12}$\,cm). Also, the nondetection of
a thermal component in the spectrum also suggests that the
photosphere component is suppressed, suggesting that the
outflow is Poynting flux dominated (e.g., Zhang \& Pe'er 2009;
Gao \& Zhang 2015).

The rapid decay of X-ray emission at the end of prompt emission
($t^{-(8.6\pm 1.12)}$) is also consistent with such a picture.
The standard model predicts that the decay slope cannot be
steeper than a decay index $\alpha = 2+\beta$, where $\beta$ is
the spectral index (Kumar \& Panaitescu 2000). Such a
``curvature effect'' prediction is valid if the outflow moves
with a constant Lorentz factor. Uhm \& Zhang (2015) pointed out
that the decay slope can be steeper than this prediction if the
outflow is undergoing acceleration. Applying the theory to
X-ray flares by properly correcting the zero-time effect (Liang
et al. 2006), Uhm \& Zhang (2016) suggested that the X-ray
flare emission region is undergoing rapid acceleration; see
also Jia et al. (2016) for an extended analysis of a larger
sample of X-ray flares. For GRB 110731A, the rapid decay of
X-ray emission at the end of prompt emission is $\alpha \approx
8.6$. However, this decay slope is dependent on the zero-time
($T_{\rm z}$), which is uncertain.

Owing to the short duration of the prompt emission, we reanalyze
the temporal behavior of the X-ray emission at the end of the prompt
emission; one has $\alpha \approx 4$ if $T_{\rm z}=4$\,s is
adopted. On the other hand, the photon index of spectra
($\Gamma_{\rm ph}$) at the end of the prompt emission is about 1.5
by invoking the power-law model fit, so $\beta= \Gamma_{\rm ph}-1
\approx 0.5$. Within the curvature-effect scenario, the temporal
index $\alpha$ and spectral index $\beta$ should satisfy
$\alpha= 2+\beta$. However, $\alpha > 2+\beta = 2.5$ if
$T_{\rm z}=4$\,s is adopted, so the curvature effect is
unlikely. Instead, the data seem to be consistent with
a model that invokes dissipation of a moderately high $\sigma$
Poynting flux in the emission region (e.g., ICMART
model; Zhang \& Yan 2011). Alternatively, if $T_{\rm z}=5$\,s
is adopted, then $\alpha\approx2.8$, only a little larger than
$2+\beta = 2.5$. In this case, the curvature effect cannot be
ruled out.

To summarize, the constraint on the emission region $R_\gamma$,
the nondetection of a thermal component in the spectrum, and
the possibility of bulk acceleration in the emission region all
point toward a consistent picture regarding the jet composition
of GRB 110731A: it is very likely Poynting flux dominated.

\section{Central Engine}

Two types of GRB central engine models have been discussed in
the literature (see, e.g., Kumar \& Zhang 2015, for a review). One
type invokes a hyperaccreting stellar-mass black hole (e.g.,
Popham et al. 1999; Narayan et al. 2001; Lei et al. 2013). The
second type invokes a rapidly spinning, strongly magnetized NS
called a millisecond magnetar, which has been invoked to
interpret the shallow-decay, long-lasting, early-afterglow
phase (Dai \& Lu 1998; Zhang \& Meszaros 2001) in both long and
short GRBs (Fan \& Xu 2006; Troja et al. 2007; Rowlinson et al.
2010, 2013; L\"{u} et al. 2015). Within the black hole central
engine, the plateau and subsequent steep decay are more
difficult to interpret. Here, we test whether the magnetar
central engine can power GRB 110731A based on the observed
properties of GRB 110731A --- i.e., the plateau phase produced
by energy injection from a magnetar wind, and the sharp drop
thereafter being due to the collapse of the magnetar forming a
black hole.

According to Zhang \& M\'esz\'aros (2001), the energy reservoir
is the total rotation energy of the millisecond magnetar,
\begin{equation}
\label{Erotaion}
E_{\rm rot} = \frac{1}{2} I \Omega_{0}^{2}
\approx 2 \times 10^{52}~{\rm erg}~
M_{1.4} R_6^2 P_{0,-3}^{-2},
\label{Erot}
\end{equation}
where $I$ is the moment of inertia, $\Omega_0 = 2\pi/P_0$ is
the initial angular frequency of the NS, $M_{1.4} = M/1.4~{\rm
M}_\odot$, and the convention $Q = 10^x Q_x$ is adopted in cgs
units for all other parameters throughout the paper. The
characteristic spin-down luminosity and spin-down timescale are
related to the magnetar initial parameters
\begin{equation}
 L_0 = (1.0 \times 10^{49})\,(B_{p,15}^2 P_{0,-3}^{-4} R_6^6)~{\rm erg~s^{-1}},
\label{L0}
\end{equation}
\begin{equation}
 \tau = (2.05 \times 10^3)\,(I_{45} B_{p,15}^{-2} P_{0,-3}^2 R_6^{-6})~{\rm s},
\label{tau}
\end{equation}
where $B_{\rm p}$ and $P_0$ correspond to the surface polar cap
magnetic field and initial spin period, respectively.

Using Eq.~\ref{L0} and Eq.~\ref{tau}, one can derive $B_{\rm
p}$ and $P_0$ as
\begin{eqnarray}
 B_{p,15} = 2.05\,(I_{45} R_6^{-3} L_{0,49}^{-1/2} \tau_{3}^{-1})~{\rm G},
\label{Bp}
\end{eqnarray}
\begin{eqnarray}
P_{0,-3} = 1.42\,(I_{45}^{1/2} L_{0,49}^{-1/2} \tau_{3}^{-1/2})~{\rm s}.
\label{P0}
\end{eqnarray}
Through light-curve fitting, one can derive the break-time
luminosity as
\begin{eqnarray}
L_b = 4\pi D^2 F_b,
\end{eqnarray}
where $F_{\rm b}$ is flux at break time $t_{\rm b}$. For a plateau, the
characteristic spin-down luminosity can be estimated as
\begin{eqnarray}
L_0 \approx L_b.
\label{L0=Lb}
\end{eqnarray}

\begin{figure}
\center
\includegraphics[angle=0,scale=0.4]{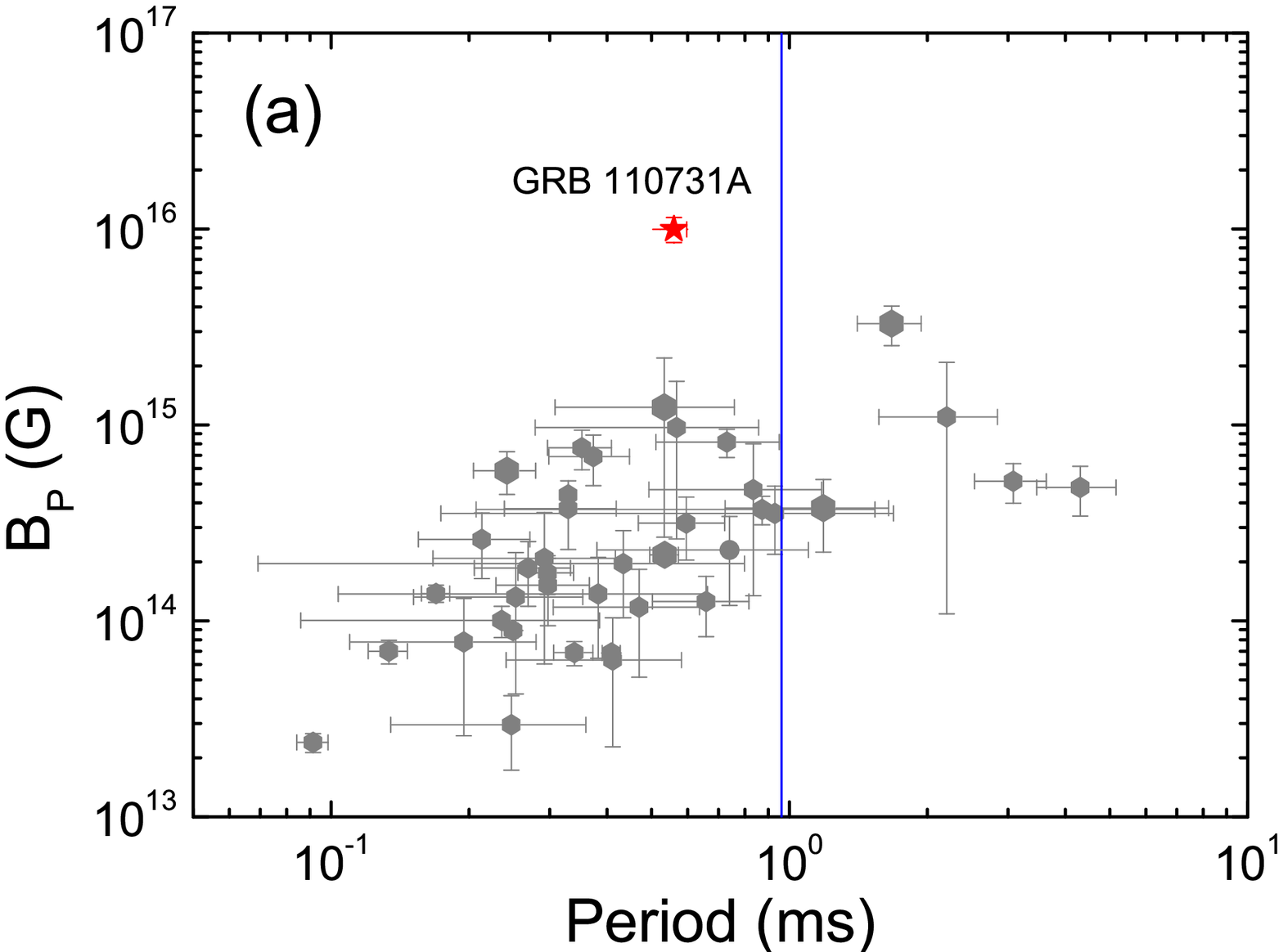}
\includegraphics[angle=0,scale=0.4]{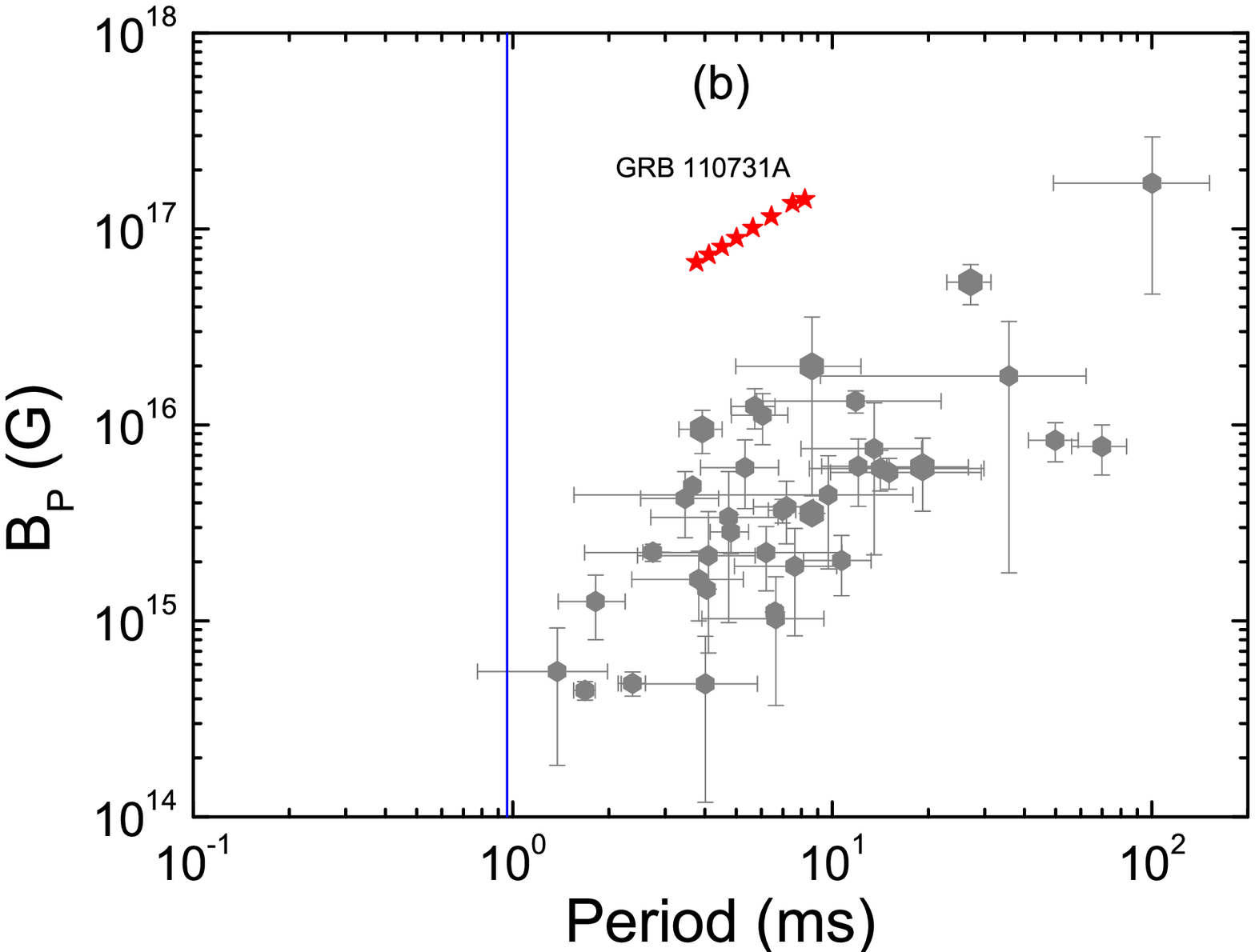}
\caption{Inferred magnetar parameters, initial spin period $P_0$ vs. surface polar cap magnetic field strength $B_p$,
of GRB 110731A (red star). (a)the case of isotropic winds; (b) The case of beaming
correction with jet opening angle
$\theta_{\rm j}$ in the range 5.5$^\circ$--12.2$^\circ$.
The vertical solid line is the breakup
spin period for an NS (Lattimer \& Prakash 2004).}
\label{fig:magnetar}
\end{figure}

The spin-down timescale can be generally identified as
\begin{eqnarray}
\tau \geq t_b/(1+z).
\label{tautb}
\end{eqnarray}
The ``greater than'' sign takes into account that the
supramassive magnetar collapses to a black hole before it is
significantly spun down. One can therefore derive $B_{\rm p} \leq
9.9\times 10^{15}$\,G and $P_0 \leq 0.56$\,ms if we assume that
the magnetar wind is isotropic (see Fig. \ref{fig:magnetar}a).
Since this value of $P_0$ is shorter than the breakup limit of
an NS ($P_0=0.96$\,ms; Lattimer \& Prakash 2004), we also
consider a possible beaming factor ($f_{\rm b}$) of the GRB outflow,
\begin{eqnarray}
\label{fb}
f_b = 1-\cos \theta_j \approx (1/2) \theta_j^2,
\end{eqnarray}
where $\theta_{\rm j}$ is jet opening angle. The jet opening angle
can be estimated as
\begin{eqnarray}
\label{theta_j}
\theta_j  &=&  0.07\,\left(\frac{t_j}{1~{\rm
day}}\right)^{3/8}\left(\frac{1+z}{2}\right)^{-3/8} \nonumber  \\& \times &
\left(\frac{E_{\rm K,iso}}{10^{53}~{\rm erg}}\right)^{-1/8}\left(\frac{n}{0.1 ~\rm
cm^{-3}}\right)^{-1/8}~{\rm rad},
\end{eqnarray}
where $n \approx 0.1$\,cm$^{-3}$ is the ambient-medium density
and $E_{\rm K,iso}$ is the kinetic energy of the outflow. The
temporal index of the normal decay segment is
$\alpha_2=1.18\pm0.01$, and the spectral index is
$\beta_X=0.85\pm0.13$. They satisfy $2\alpha_2 \approx
3\beta_X$, suggesting a spectral regime $\nu_m < \nu < \nu_c$.
Following Zhang et al. (2007), we derive
\begin{eqnarray}
E_{\rm K,iso,52} & = & \left[\frac{\nu F_\nu (\nu=10^{18}~{\rm Hz})}{6.5\times
10^{-13} ~{\rm erg\,s^{-1}\,cm^{-2}} }\right]^{4/(p+3)} \nonumber  \\& \times &
D_{28}^{8/(p+3)}(1+z)^{-1}
 t_d^{3(p-1)/(p+3)}\nonumber \\
& \times &f_p^{-4/(p+3)} \epsilon_{B,-2}^{-(p+1)/(p+3)}
\epsilon_{e,-1}^{4(1-p)/(p+3)} \nonumber \\ & \times & n^{-2/(p+3)}
\nu_{18}{^{2(p-3)/(p+3)}},
\nonumber \\
\end{eqnarray}
where $p=2\beta+1$. With standard values of microphysics
parameters (e.g., Panaitescu \& Kumar 2002; Yost et al. 2003),
such as $\epsilon_{\rm e} = 0.1$, $\epsilon_B = 0.01$, and $n
\approx 0.1$\,cm$^{-3}$, we derive $E_{\rm K,iso} \approx
7.5\times10^{52}$\,erg. No jet break feature was detected up to
$\sim 7.5$\,days of afterglow emission; we therefore set a
lower limit on the jet opening angle, $\theta_{\rm
j}>5.5^{\circ}$.

Within the magnetar model, one can also set an upper limit of
$\theta_j$ by requiring that the total energy does not exceed
the total spin energy of the magnetar: $E_{\rm rot}\geq
E_\gamma+E_{\rm K}+E_{\rm flare}$, where $E_\gamma=E_{\gamma,
{\rm iso}} f_b$, $E_{\rm K} = E_{\rm K, iso} f_b$, and $E_{\rm
flare} = E_{\rm flare, iso} f_b$ is the energy of the GeV
flare. This gives $\theta_{\rm j} < 12.2^{\circ}$. By including
the range of the opening angle $5.5^{\circ}< \theta_{\rm j} <
12.2^{\circ}$, we rederive the magnetar parameters for GRB
110731A (red stars in Fig. \ref{fig:magnetar}b). We find that
the $P_0$ upper limit is in the allowed range (longer than the
breakup limit), but the inferred $B_{\rm p}$ upper limit is
significantly larger than the $B_{\rm p}$ values inferred for
other GRBs. In any case, since $B_{\rm p}$ is an upper limit,
the magnetar model is possible given that the spin-down time
$\tau$ is much longer than $t_{\rm b}/(1+z)$.

Within this supramassive NS collapsing framework, there is
another energy budget, the total magnetic field energy after
the magnetosphere is expelled following the collapse of the NS.
It may be estimated as (Zhang 2014)
\begin{eqnarray}
E_{\rm B, iso} &\approx& \int^{R_{\rm LC}}_{R}4\pi r^2
\frac{B^2_p}{8\pi}\left(\frac{r}{R}\right)^{-6} dr \nonumber \\& \approx &
(1/6)B^2_p R^3 \approx (1.7\times 10^{47})\, B^2_{p, 15} R^3_{6}~ {\rm erg},
\label{EB}
\end{eqnarray}
where $R_{\rm LC}\gg R$ is the light-cylinder radius. This is a
relatively small energy for typical parameters, but can be
important if $B_{\rm p}$ is large (close to the upper limit inferred
above).

\begin{figure}
\center
\includegraphics[angle=0,scale=0.4]{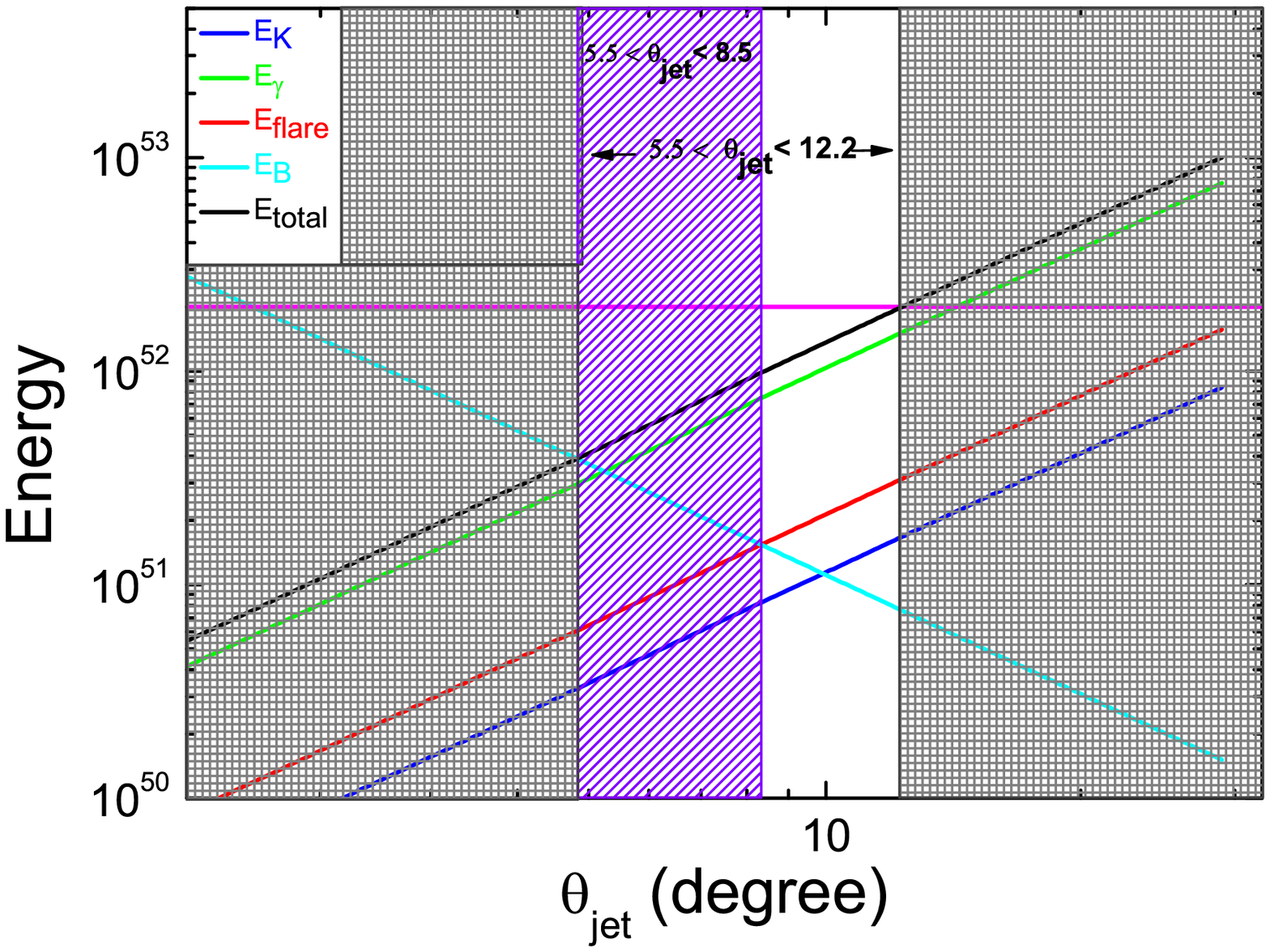}
\includegraphics[angle=0,scale=0.4]{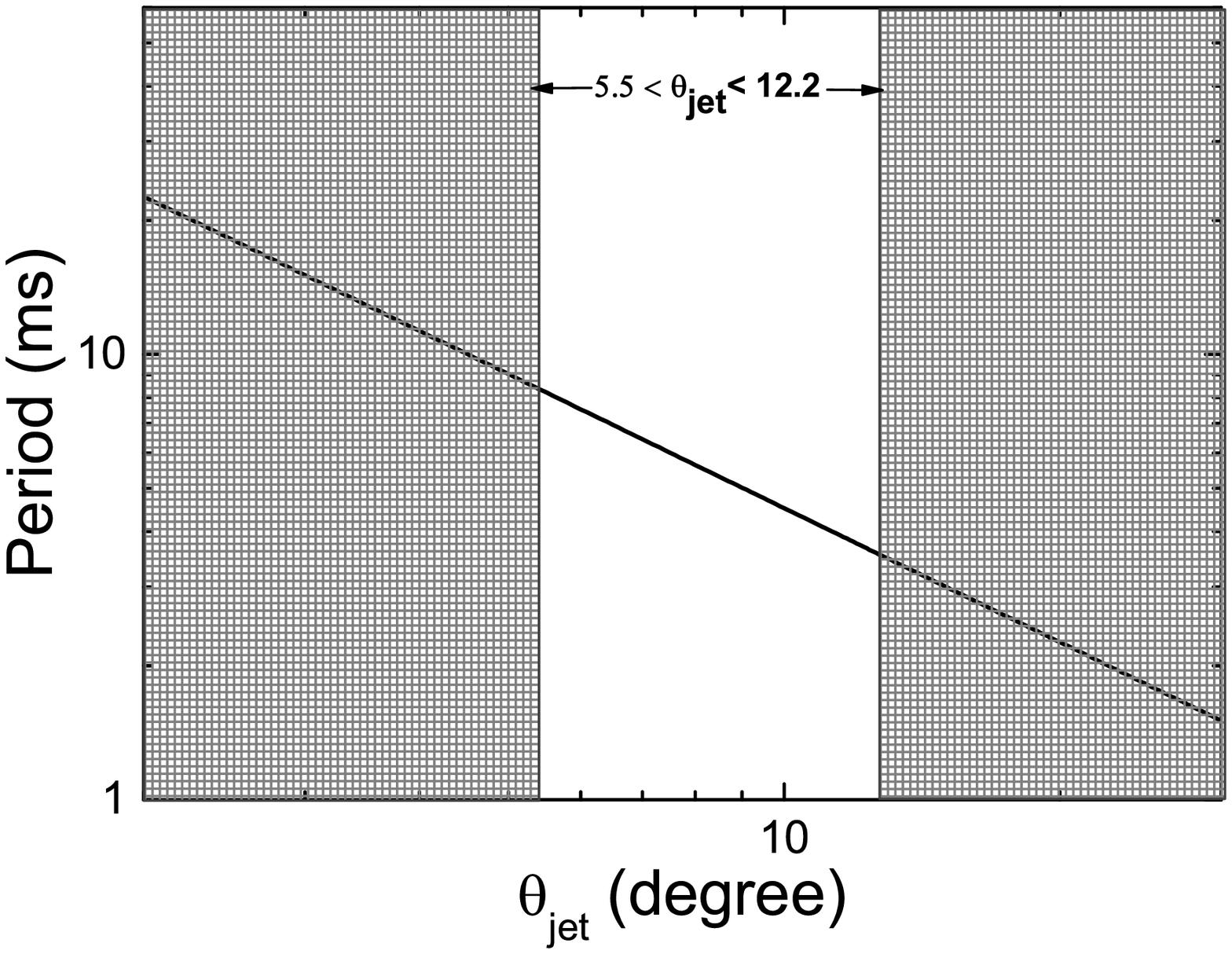}
\caption{(a)Energy $E_{\gamma}$ (green line), $E_{\rm K}$ (blue line), $E_{\rm
flare}$ (red line),
$E_{\rm B}$ (cyan line), and $E_{\rm total}$ (black line) as a function of jet opening
angle. The horizontal line is
the total energy budget of the magnetar ($E_{\rm rot}\approx 2\times 10^{52}$\,erg);
(b)Inferred upper limit of $P_0$ as a function of $\theta_{\rm j}$.}
\label{fig:energy}
\end{figure}

Figure~\ref{fig:energy} gives all the energy components as a
function of jet opening angle, with the allowed $\theta_{\rm
j}$ range marked.

\section{Progenitors}

With the measured $z = 2.83$, GRB 110731A has a rest-frame
duration $[T_{90}/(1 + z)]$ shorter than 2\,s. This naturally
raises the interesting question regarding the progenitor system
of this burst (Type I vs. Type II; Zhang et al. 2009). In the
past, there have been quite a few GRBs that are rest-frame
short, including several high-$z$ GRBs such as GRB 080913 at
$z=6.7$ with $T_{90} =8$\,s (Greiner et al. 2009), GRB 090423
at $z = 8.2$ with $T_{90} = 10.3$\,s (Salvaterra et al. 2009;
Tanvir et al. 2009), and GRB 090429B at $z = 9.4$ with $T_{90}
= 5.5$\,s (Cucchiara et al. 2011). Multiwavelength observed
properties suggested that those three high-$z$ GRBs are likely
of a Type II (massive star) origin (Zhang et al. 2009).

\begin{figure}
\center
\includegraphics[angle=0,scale=0.4]{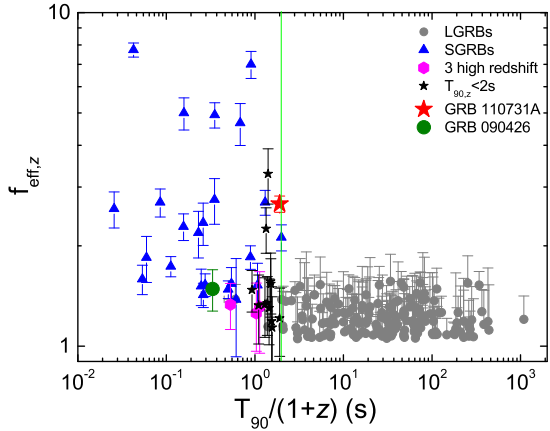}
\caption{$T_{90}/(1 + z)-f_{{\rm eff},z}$ diagram of
both long and short GRBs taken from L\"{u} et al (2014). The red
star denotes GRB 110731A, and the vertical line is $T_{90}/(1+z)=2$\,s.}
\label{fig:ft90}
\end{figure}

L\"{u} et al. (2014) proposed a method to judge whether a
rest-frame short GRB is more likely the ``tip of iceberg'' of a
long-duration GRB. They move a specific long GRB to
progressively larger redshifts until the rest-frame duration is
shorter than 2\,s and then define the ratio of the peak flux of
this pseudo-GRB to the background flux as $f_{{\rm eff},z}$.
The $f_{{\rm eff},z}$ value of long GRBs is typically smaller
than 2. The three high-$z$ GRBs all have $f_{{\rm eff},z}$
smaller than 2, suggesting that they are consistent with being
a long GRB as observed at high redshift. We perform the same
analysis for GRB 110731A following L\"{u} et al. (2014), and
find that the value is $f_{{\rm eff},z}=2.67$. As shown in
Figure \ref{fig:ft90}, this value (red star) is higher than
that of typical long GRBs (gray), but is consistent with short
GRBs (blue triangles). Following L\"{u} et al. (2014), we also
calculate the probability ($p$) of this being a disguised short
GRB: $p\approx 0.03$. In general, these arguments suggest that
the GRB is consistent with having a Type I (compact-star
merger) origin.

\begin{figure}
\center
\includegraphics[angle=0,scale=0.6]{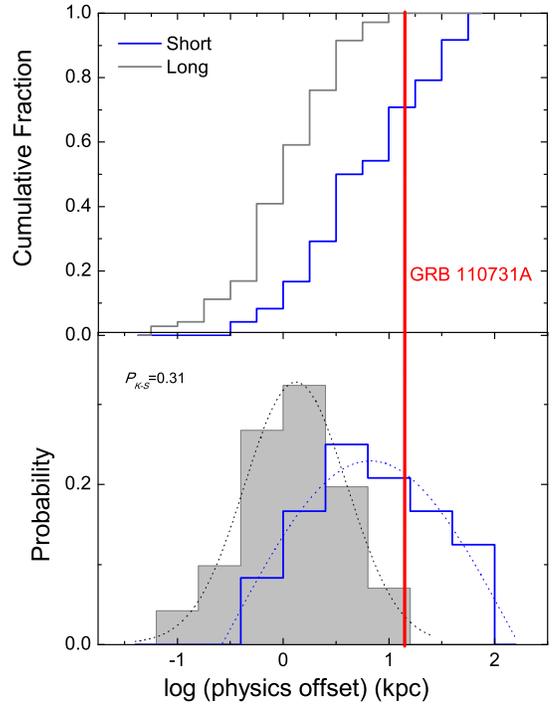}
\hfill\caption{Bottom panel: the distribution of physical offsets
for short GRBs (blue; Fong et al. 2010) and long GRBs (gray; Blanchard et al.
2016); the value of the K-S test is about 0.31. The blue dotted line and
gray dotted line indicate the best Gaussian fits. Top panel: a
cumulative distribution of long (gray) and short (blue) GRBs. The red
vertical line corresponds to GRB 110731A.}
\label{fig:offset}
\end{figure}

The host-galaxy information seems to also support a Type I
origin for GRB 110731A. As discussed in \S~2.5, our Keck
observation revealed an extended source northeast of the
afterglow position, which might be the host galaxy of GRB
110731A. The source has an $I$-band magnitude $m_I \approx
23.5$. The offset between this source and the GRB afterglow is
$1.63''$, which corresponds to $\sim13.0$\,kpc, with the
possibility of a physical association being $\sim 3$\%. First,
we compare the distribution of long GRB physical offsets with
those of short GRBs; the K-S test yields $P_{\rm K-S}=0.31$,
suggesting that the objects are not drawn from the same
population. Then, we compare the offset of GRB 110721A with the
distributions of the physical offsets of both long and short
GRBs observed by the {\em Hubble Space Telescope} (Fong et al.
2010; Blanchard et al. 2016). We find that the physical offset
of GRB 110731A is larger than that of almost all long GRBs, but
is statistically consistent with typical short GRBs (see Fig.
\ref{fig:offset}). This also supports the Type I origin of the
burst, if the extended source is indeed the host galaxy of GRB
110731A.

\begin{figure}
\center
\includegraphics[angle=0,scale=0.3]{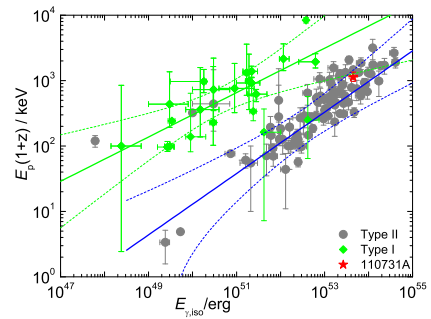}
\caption {$E_{\rm p}$-$E_{\rm \gamma,iso}$ diagram of both long (blue dots) and short
(green diamonds) GRBs. The data points for long and
short GRBs are taken from Amati et al. (2002) and Zhang et al. (2009),
respectively.
The solid lines are the best fit with a power-law model, and
dashed lines mark 3$\sigma$ confidence bands. The red star
corresponds to GRB 110731A.}
\label{fig:Amati}
\end{figure}

Empirically, most long GRBs are found to satisfy a relationship
between $E_{\rm p}(1+z)$ and $E_{\gamma, {\rm iso}}$ ($E_{\rm
p}(1+z)\propto E^{1/2}_{\gamma, {\rm iso}}$; Amati et al.
2002), but outliers do exist (e.g., long GRBs 980425, 031203,
and 050826). In contrast, most short GRBs are inconsistent with
this empirical relation (Zhang et al. 2009) and seem to follow
a different correlation with a larger dispersion. However,
outliers to this relation also exist (e.g., short GRB 060121).
In order to compare GRB 110731A with long and short GRBs, we
calculate $E_{\gamma, {\rm iso}}\approx 4.5\times 10^{53}$\,erg
from 1\,keV to $10^{4}$\,keV based on the spectral properties
and then plot it in the $E_{\rm p}(1+z)$--$E_{\gamma, {\rm
iso}}$ diagram (see Fig. \ref{fig:Amati}). GRB 110731A falls in
the 3$\sigma$ confidence band of power-law fitting of long
GRBs. However, this empirical relation does not provide
persuasive evidence that GRB 110731A is from massive star core
collapse (outliers do exist), and it may be caused by some
selection effects.


\section{The Origin of the GeV Flare}

The origin of the GeV flare is mysterious. As shown in Figure
3, it rises during the prompt emission phase but peaks after
the BAT-band emission finished. The time-integrated spectral
energy distribution is fitted by a Band function alone,
and it seems that the GeV emission has the same origin.
However, when we analyze the time-resolved spectra during the
GeV flare (from $T_0+7.3$\,s to $T_0+8.6$\,s), they are well
fitted by a Band function with an additional power-law
component (Table 1 and Fig. \ref{fig:GBMSP}). This suggests
that the GeV flare may have a distinct origin from the sub-MeV
emission.

If this is the case, one possible scenario is to interpret the
GeV flare within the framework of a supramassive NS collapsing
into a black hole. Zhang (2014) suggested that the ejection of
the magnetosphere may be accompanied by a fast radio burst. It
is possible that such an ejection may power a GeV flare.
However, the total amount of magnetospheric energy (Eq.~17) is
typically smaller than the GeV flare energy, so one runs into
an energy budget problem. Nonetheless, the uncertainty in the
beaming factor $f_{\rm b}$ makes it possible that $E_{\rm B}> E_{\rm
flare}$ in a certain range of jet opening angles. From Figure
9a, one can derive that the condition is $5.5^{\circ}<\theta_{\rm j}
< 8.5^{\circ}$. The specific emission mechanism depends on the
particle acceleration details within the ejected magnetosphere,
but the sudden acceleration of the magnetosphere makes it
plausible to have a GeV peak right after the rapid decline in
sub-MeV emission.

Alternatively, the GeV emission may be produced from the
external shock (e.g., Kumar \& Barniol-Duran 2009, 2010;
Ghisellini et al. 2010; Zhang 2011; Maxham et al. 2011; He et
al. 2012; Liu et al. 2012). The initial steep decay may suggest
a reverse-shock component (e.g., Wang et al. 2001, 2002).
However, this interpretation is in conflict with our suggestion
of the X-ray peak at 65\,s as the deceleration time.

\section{Conclusions and Discussion}

GRB 110731A is a peculiar GRB with a duration of $\sim7.3$\,s
detected by {\em Swift} and {\em Fermi}, and a measured
redshift of $z=2.83$. The total isotropic-equivalent energy in the
10\,keV--10\,GeV range is $E_{\gamma, {\rm iso}}\approx (6.8\pm
0.1)\times 10^{53}$\,erg for the prompt emission. One GeV flare
was detected by LAT with its highest photon energy $\sim
2$\,GeV, and the peak time of the GeV flare corresponds to the
transition break time from the prompt emission (plateau) to a
steeper decay. The total isotropic-equivalent energy of the GeV
flare is $E_{\rm iso, flare}\approx 1.4\times 10^{53}$\,erg.
Our Keck $I$-band image of the field placed an upper limit to
the host-galaxy brightness ($m_I \approx 24.9$\,mag) at the
afterglow position and identified a putative galaxy with a
physical offset of $\sim13$\,kpc. We analyze the broadband
data and compare them with GRB theoretical models, finding the
following results.
\begin{itemize}
 \item We apply both the pair-opacity method and afterglow
     deceleration time method to constrain the Lorentz
     factor of the ejecta. The former gives $\Gamma>190$,
     while the latter gives $\Gamma \approx 580$ and
     $\Gamma\approx 154$ within the homogeneous and wind
     density profiles, respectively.
 \item The broadband featureless Band-function spectra
     cover 5--6 orders of magnitude in energy, as well as
     the very high energy gamma-ray emission ($\geq
     1$\,GeV). Nondetection of a thermal component may be
     consistent with a Poynting-flux-dominated flow as the
     jet composition of the burst. On the other hand,
     during the rapid decay phase following prompt
     emission, the temporal decay index ($\alpha$) is
     steeper than the curvature-effect prediction
     $\alpha=2+\beta$, which supports possible acceleration
     of the emission region. However, this decay slope is
     dependent on the selected zero-time, which is
     uncertain.
 \item The central engine of the GRB may be a millisecond
     magnetar, but with a relatively large upper limit of
     both $B_p$ and $P_0$. With a beaming correction, the
     total observed energy (sum of $E_{\rm \gamma}$,
     $E_{\rm K}$, and $E_{\rm flare}$) is within the energy
     budget provided by the spin energy of the millisecond
     magnetar ($E_{\rm rot}\approx 2\times 10^{52}$\,erg)
     when the jet opening angle satisfies $5.5^{\circ}<
     \theta_{\rm j} < 12.2^{\circ}$.
 \item The burst has a rest-frame duration shorter than
     2\,s. A relatively large $f_{{\rm eff},z}$ value and a
     substantial physical offset from the putative host
     galaxy suggest that the progenitor of GRB 110731A is
     likely to come from a compact-star merger.
 \item The GeV flare is mysterious. However, within the
     magnetar central scenario, the GeV flare may be
     produced during the ejection of the magnetosphere when
     the magnetar collapses to form a black hole.
\end{itemize}


\acknowledgments

We acknowledge the use of public data from the {\it Swift} and
{\it Fermi} data archives and the UK {\it Swift} Science Data
Center. We thank Xue-Feng Wu and Wei-Hua Lei for helpful
comments and discussions, as well as D. Alexander Kann for
observation suggestions. This work is supported by the National
Basic Research Program (973 Programme) of China 2014CB845800,
the National Natural Science Foundation of China (grant nos.
11603006, 11533003, 11673006, 11603003, 11543005, U1331202,
11303005, 11363002), the One-Hundred-Talents Program of Guangxi
colleges, the high-level innovation team and outstanding
scholar program in Guangxi colleges, the Guangxi Science
Foundation (2016GXNSFCB380005, 2016GXNSFFA380006,
2013GXNSFFA019001, 2014GXNSFAA118011), and the Scientific
Research Foundation of Guangxi University (grant no.
XGZ150299). A.V.F.'s group at UC Berkeley has been supported
by Gary and Cynthia Bengier, the Richard \& Rhoda Goldman Fund,
the Christopher R. Redlich Fund, the TABASGO Foundation, and US
NSF grant AST-1211916. Some of the data presented herein were
obtained at the W. M. Keck Observatory, which is operated as a
scientific partnership among the California Institute of
Technology, the University of California, and NASA; the
observatory was made possible by the generous financial support
of the W. M. Keck Foundation. Research at Lick Observatory is
partially supported by a generous gift from Google.

\begin{center}
\begin{deluxetable}{lllllllllllll}
\tablewidth{0pt} \tabletypesize{\footnotesize}
\tabletypesize{\tiny} \tablecaption{Fitting Parameters of the
GBM-LAT Spectrum of GRB 110731A.}\tablenum{1}

\tablehead{\colhead{Interval (s)}& \colhead{}& \colhead{Band}&
\colhead{}& \colhead{Power-law}&  \colhead{PGSTAT/dof}}

\startdata
& $\hat{\alpha}$   &$\hat{\beta}$    &$E_{\rm p}(keV)$	&	$\lambda\tablenotemark{a}$ & $\chi^2$/dof	\\
\hline 	
$0--8.6$                 & $-0.89\pm 0.06$  &$-2.32\pm0.03$  &$285\pm 41$	&	-- &	373/315	\\
\hline 	
$0--2\tablenotemark{b}$  & $-0.87\pm 0.12$  &--              &$145\pm 21$	&	-- &	228/242	\\
$2--7.3$                 & $-0.74\pm 0.09$  &$-2.32\pm 0.03$ &$277\pm 59$	&	-- &	254/315	\\
$7.3--8.6$               & $-1.08\pm 0.29$  &$-2.15\pm 0.16$ &$2793\pm 673$	&	$1.57\pm 0.47$		&	263/321	\\
																	
\enddata
\tablenotetext{a}{The power-law index in the $\nu F_{\nu}$
spectrum.}
\tablenotetext{b}{In this time interval, the spectrum
is fitted well by a cutoff power-law model owing to a lack of high-energy
photons.}

\end{deluxetable}
\end{center}



\end{document}